\documentclass[twocolumn,secnumarabic,amssymb, nobibnotes, aps, prd]{revtex4}
\usepackage{graphicx}

\begin{document}
\title{Energy of magnetic moment of superconducting current in magnetic field }
\author{V.L. Gurtovoi  and  A.V. Nikulov}
\affiliation{Institute of Microelectronics Technology and High Purity Materials, Russian Academy of Sciences, 142432 Chernogolovka, Moscow District, RUSSIA.} 
\begin{abstract} The energy of magnetic moment of the persistent current circulating in superconducting loop in an externally produced magnetic field is not taken into account in the theory of quantization effects because of identification of the Hamiltonian with the energy. This identification misleads if, in accordance with the conservation law, the energy of a state is the energy expended for its creation. The energy of magnetic moment is deduced from a creation history of the current state in magnetic field both in the classical and quantum case. But taking this energy into account demolishes the agreement between theory and experiment. Impartial consideration of this problem discovers the contradiction both in theory and experiment. 
\end{abstract}

\maketitle 

\narrowtext

\section{Introduction}
It is well known \cite{FeynmanLec2} that electric current $I$ circulating clockwise or anticlockwise in a flat loop with a vector area $\bf{S}$ induces magnetic dipole moment equal ${\bf M _{m}} = I\bf{S}$. It is well known also \cite{FeynmanLec2} that magnetic moment in an externally produced magnetic field $\bf{B}$ has an energy equal  $E _{M} = - \bf{M _{m}}\bf{B}$. But this energy is not taken into account in the theory describing quantization effects in superconducting loop \cite{Tinkham}.  This discrepancy is particularly demonstrable in the case of persistent-current qubits \cite{Mooij99} or flux qubits \cite{Mooij03}. Flux qubits consist of a superconducting loop interrupted by either one or three Josephson junctions \cite{Clarke08}. The two quantum states of flux gubit are persistent current $I _{p}$ circulating in the loop clockwise and anticlockwise in an externally produced magnetic field $\bf{B}$ corresponding approximately the half ${\bf BS}= \Phi \approx (n+0.5) \Phi _{0}$ of the flux quantum $\Phi _{0} = \pi \hbar /e$ inside the loop \cite{Clarke08}. The qubit effective Hamiltonian are represented by the Pauli spin matrices $\sigma _{z}$, $\sigma _{x}$ \cite{Mooij03,MakhlinRMP}, that is
$$H_{q} = \epsilon \sigma _{z} - \Delta \sigma _{x} \eqno{(1)}$$
as well as the Hamiltonian of spin - 1/2. The energy difference between two spin states of electron, for example, is the energy $\epsilon  = \mu _{B}B_{z}$ of magnetic moment equal the Bohr magneton $\mu _{B} =  -e\hbar /2m$ in external magnetic field $B_{z}$. This energy of flux qubit should be equal $|E _{M}| = M _{m}B _{z} = I _{p}\Phi $ when ${\bf B} = (0,0,B _{z})$ and the flux qubit loop is in the flat $x - y$. But this energy is not take into account although the energy considered in the theory \cite{Mooij03} $\epsilon  =I_{pm}\Phi _{0}(\Phi /\Phi _{0}  - 1/2)$ (where $I _{pm}$ is the maximum qubit persistent current) is much lower than the energy $|E _{M}| = M _{m}B _{z} = |I _{p}\Phi | \approx |I _{p}|\Phi _{0}/2 $ near the half of the flux quantum $|\Phi /\Phi _{0}  - 1/2| \ll 1$. 

\section{Quantization effects in superconductors} 
 \label{}
The two states of flux qubit are assumed at $\Phi \approx (n+0.5) \Phi _{0}$ because of the requirement $\oint _{l} dl \nabla \varphi  =  2\pi n$ that the complex pair wave function $\Psi = |\Psi |e^{i\varphi }$ must be single - valued at any point in a superconductor. 

\subsection{Quantization of angular momentum}
Superconducting loop without Josephson junctions should also have such two states due to this requirement or the quantization of angular momentum of Cooper pair 
$$m _{p} = \oint _{l} dl p/2\pi = \oint _{l} dl \hbar \nabla \varphi /2\pi =  \hbar n \eqno{(2)}$$ 
The idea of flux qubit presupposes the superposition of two macroscopic quantum states \cite{Clarke08} assumed first by Anthony Leggett in the 1980s \cite{Leggett1987}. We will not consider the problem of superposition (described with the term $\Delta \sigma _{x} $ in (1)) assumed only in a superconducting loop interrupted by Josephson junctions. We take an interest in the energy difference $\epsilon $ of the term $\epsilon \sigma _{z}$ between two permitted states. Therefore superconducting loop without Josephson junctions will be considered first of all.

The relation 
$$\mu _{0}\oint_{l}dl \lambda _{L}^{2} j  + \Phi = n\Phi_{0}  \eqno{(3)}$$ 
deduced from the requirement (2) can describe the Meissner effect, magnetic flux quantization and quantization of pair velocity or persistent current \cite{FPP2008}. The Meissner effect i.e. the expulsion of magnetic flux $\Phi $ from the interior of a superconductor, discovered by Meissner and Ochsenfeld in 1933 \cite{Meissner}, is observed in a bulk entire superconductor in which the wave function $\Psi = |\Psi |e^{i\varphi }$ has no singularity and therefore the quantum number $n = 0$ and, according to (3), $\Phi = n\Phi _{0} = 0$ inside superconductor where the density of superconducting current $j = 0$. The flux quantization was observed first in 1961 \cite{FQ1961} with the help of measurements of magnetic flux trapped in hollow superconducting cylinder the wall width $w$ of which is larger $ w \gg \lambda _{L}$ than the London penetration depth $\lambda _{L} = (m/\mu _{0}q^{2}n_{s})^{0.5} = \lambda _{L}(0)(1 - T/T_{c})^{-1/2}$ \cite{London35} ($\lambda _{L}(0) \approx 50 \ nm = 5 \ 10^{-8} \ m$ for most superconductors \cite{Tinkham}). The current density $j \approx  0$ along a contour $l$ inside superconducting region in this case of strong screening and therefore $\Phi \approx  n\Phi _{0}$ according to (3). 

The quantization of the persistent current  
$$\frac{\lambda _{L}^{2}}{s} \mu _{0}lI _{p} = (n\Phi_{0} - \Phi ) \eqno{(4)}$$ 
or velocity $\oint_{l}dl v  =  (2\pi \hbar /m)(n - \Phi /\Phi_{0}) $ is observed in the case of weak screening, for example in a loop with section $s \ll \lambda _{L}^{2}$. The kinetic inductance $L _{k} \approx (\lambda _{L}^{2}/s) \mu _{0}l$ exceeds in this case the magnetic inductance $L  \approx \mu _{0}l$ and one can always neglect the magnetic flux $\Delta \Phi _{I} = L I _{p}$ induced with the current $I _{p}$ for a sufficiently thin superconductor \cite{Tinkham}. Therefore the magnetic flux $\Phi = BS + L I _{p}$ equals approximately the one $\Phi \approx BS $ of externally produced magnetic field $B$. Quantization effect in the weak screening limit was observed first by W. A. Little and R. D. Parks \cite{LP1962} at measurements of the resistance of thin cylinder in the temperature region corresponding to its superconducting resistive transition. Later on quantum periodicity of other quantities were observed: ring resistance \cite{Letter07,toKulik2010}, magnetic susceptibility \cite{PCScien07}, critical current \cite{JETP07J} and dc voltage measured on segments of asymmetric rings \cite{Letter07,toKulik2010,PerMob2001,Letter2003,PCJETP07,PLA2012}. Superconducting ring according to (4), as well as flux qubit, has at $\Phi = (n'+0.5)\Phi _{0}$ the two permitted current states $I _{pm} = (n\Phi_{0} - \Phi )/ L _{k} = - 0.5\Phi_{0}/ L _{k}$ when $n = n'$ and $I _{pm} = +0.5\Phi_{0}/ L _{k}$ when $n = n'+1$. 

\begin{figure}
\includegraphics{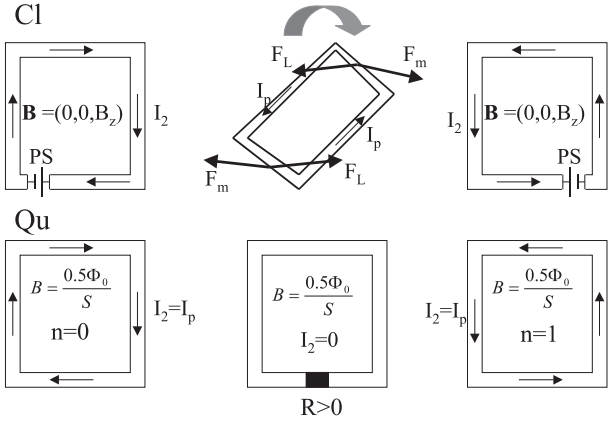}
\caption{\label{fig:epsart} Cl: The electric current $I _{2}$ circulating in the loop induces magnetic dipole moment ${\bf M _{m}} = I _{2}\bf{S}$. Moment ${\bf \tau } = {\bf M _{m}} \times {\bf B}$ of force $F _{L}$ acts on this loop in magnetic field ${\bf B}$. The moment of mechanical force $F _{m}$ should be applied and the energy $E _{M} = \int _{0}^{\pi }d\theta M _{m}B _{z}\sin\theta = 2M _{m}B _{z}$ should be expended in order to overturn the loop when the constant value of the current $I _{2}$ is maintained with the help of power source PS. Clockwise current changes into anticlockwise current relatively the $B _{z}$ direction after this turning-over. Qu: The direction of the persistent current in superconducting loop changes with quantum number $n$  change as a result of the transition of a loop segment in normal state (black) with a non-zero resistance $R > 0$ and posterior retrieval it in superconducting state.}
\end{figure}  

\subsection{Energy and Hamiltonian }  
The energy difference $\epsilon $ of these states is deduced from the Hamiltonian   
$$H = \frac{1}{2m}\sum _{a}[-i\hbar \nabla _{a} - qA(r_{a})]^{2} +U   \eqno{(5)}$$
used for description of quantization effects in superconductors as far back as 1961 \cite{QuTh1961}. According to this Hamiltonian the energy of a sufficiently thin superconducting loop with homogeneous Cooper pair density $|\Psi |^{2} = n _{s}$ should equal $\int _{V}dV\Psi *H\Psi =  \int _{V}dV|\Psi |^{2}[(1/2m)(p - qA)^{2} + U] =  \int _{l}dl sn _{s} \frac{mv ^{2}}{2} + \int _{V}dVn _{s}U$. The potential energy $\int _{V}dVn _{s}U$ does not depend on magnetic flux $\Phi $ and is not considered in the theory of quantization. The kinetic energy of Cooper pairs
$$E_{k} = \oint _{l} dl sn _{s} \frac{mv ^{2}}{2} =  \frac{I_{p}}{q}\oint _{l} dl \frac{mv}{2} = L _{k}I _{p}^{2}/2 \eqno{(6)}$$
does not depend on direction of the velocity $v$ or the current $I_{p}$. Thus, two permitted state $n$ and $n+1$ with different angular momentum  have the same energy $\int _{V}dV\Psi ^{*} H\Psi = L _{k}I _{pm}^{2}/2 + \int _{V}dVU n _{s} = I _{pm}0.5\Phi _{0} /2 + \int _{V}dVU n _{s} $ at $\Phi = (n+0.5) \Phi _{0}$ and the energy difference $\epsilon = 0$ according to the canonical Hamiltonian (5). 

But we know that the energy of two states having different magnetic moment in non-zero magnetic field should be different. Clockwise electric current $I _{p}$ can be obtained from anticlockwise current $I _{p}$ with the help of the turning-over of the loop, Fig.1Cl. It is well known that we should expand the energy $E _{M} = \int _{0}^{\pi }d\theta M _{m}B _{z}\sin\theta = 2M _{m}B _{z}$ in order the rotate the magnetic dipole moment ${\bf M _{m}} = I _{pm}\bf{S}$ in magnetic field $B _{z}$ \cite{FeynmanLec2}. "But when we go over to the Hamiltonian formalism by the standard 'canonical' procedure, the total Hamiltonian $(1/2m)(p-qA)^2$ turns out to be just the kinetic energy $mv^2/2$! Where has the 'magnetic' energy gone?" \cite{Leggett2014}. 

\section{Identification of Hamiltonian with energy misleads}   
Anthony Leggett has surmised soundly that "Perhaps our naive tendency to identify the Hamiltonian with the 'energy' is (as in some cases involving time-dependent forces) misleading?" \cite{Leggett2014}. 

\subsection{Energy expended for the current in magnetic field} 
Indeed, the energy of electric current circulating in a perfect conductor deduced from the classical Hamiltonian (16.10) in \cite{LLFT}
$$H = \frac{1}{2m}(p-qA)^{2} + q \phi \eqno{(7)}$$ 
turns out to be just the kinetic energy $\int _{V}dV n _{q}H =  \int _{V}dV n _{q}(1/2m)(p - qA)^{2} =  \int _{l}dl sn _{s} mv ^{2}/2 = L _{k}I^{2}/2$  at weak screening $L \ll  L _{k}$ as well as in the quantum case (6). This energy should be expended
$$\int _{t}dtIV = \int _{t}dtI(L_{k}+L)\frac{dI}{dt} = \frac{(L_{k}+L)I_{2}^{2}}{2} \approx   \frac{L_{k}I_{2}^{2}}{2}   \eqno{(8)}$$
by a power source in order to create the current $I_{2}$ in a loop with the inductance $L \ll  L _{k}$ in zero magnetic field, Fig.2Cl. But the power source should expend an additional energy 
$$\int _{t}dtI_{2}V = I_{2}\int _{t}dt \frac{d\Phi }{dt} = I_{2}\Phi  \eqno{(9)}$$
in order to provide zero electric field $V - d\Phi /dt = 0$ and to maintain the current $I_{2}$ when the externally produced magnetic field changes from $B = 0$ to $B = \Phi /S$, Fig.2Cl. The energy (9) deduced from the history (involving time-dependent forces) of the state equals the energy $E _{M} = - \bf{M _{m}B}$ of magnetic dipole moment ${\bf M _{m}} = I_{2}\bf{S}$ in magnetic field $\bf{B}$. This energy is positive when  $ I_{2}V > 0$ and negative when $ I_{2}V < 0$ in (9). 

Thus the total energy of the current circulating in the loop with the section $s \ll \lambda _{L}^{2}$ is the sum $E _{tot} = E _{k} + E _{M}$ of the kinetic energy (8) and the energy of magnetic dipole moment in magnetic field (9). This energy can be easily deduced for any case when magnetic field $B = \Phi /S$ appears after the current $I_{2}$ because the current $I_{2}$ constant in time has no influence on the energy expended by a power source producing the magnetic field. It is obvious this energy must have the same value when the current $I_{2}$ appears after magnetic field $B = \Phi /S$. The energy $E _{M}$ is provided in this case by the power source producing the magnetic flux due to the mutual inductance. This energy can be easily calculated in few cases, for example when the mutual inductance $L _{1,2}$ between two loops equals the inductance $L _{1}$, $L _{2}$ of each of the loops. The equalities $L _{1,2} = L _{1} = L _{2}$ take place when the magnetic flux inside the loop is induced with the help of a similar loop, located side by side, Fig.2Qu.  

\begin{figure}
\includegraphics{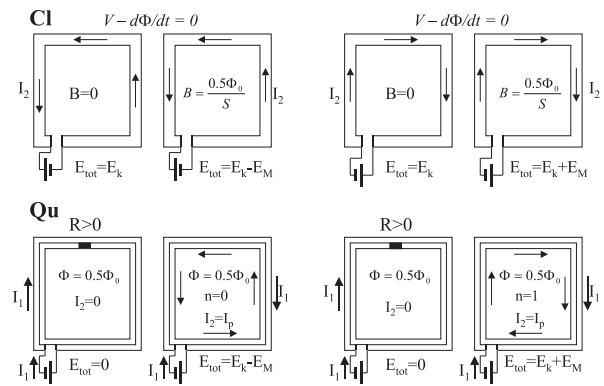}
\caption{\label{fig:epsart}  Cl: The power source should expend different energy $E _{tot} = E _{k} - E _{M}$ and $E _{tot} = E _{k} + E _{M}$ in order to create clockwise and anticlockwise current $I _{2}$ in the same magnetic field $B = 0.5\Phi _{0}/S$. Qu: The first (external) loop creates magnetic flux $\Phi = L _{1}I _{1} = 0.5\Phi _{0}$. The electric field $E = -dA/dt = - l^{-1}d\Phi /dt \approx  - l^{-1}L_{1,2} I _{p}/t _{?}$ should be induced in this loop during appearing of the persistent current $I _{2} = I _{p} \ll I _{1}$ in the second loop after transition of its normal segment (black) in superconducting state. This electric field should have opposite direction at the transition into superconducting states with different quantum number $n = 0$ and $n = 1$ because of opposite direction of the current $I _{p}$. Therefore the power source should expend different energy $- E _{M}$ and $+ E _{M}$ in order to maintain the current $I _{1}$.}
\end{figure} 

\subsection{The energy $E _{M} $ in the quantum case}
This case is important for a deduction of the $E _{M}$ energy in the quantum case because the persistent current appears after magnetic flux. The electric current exists due to the power source in the classical case shown on Fig.1Cl, Fig.2Cl and due to the quantization (2) in the quantum case shown on Fig.1Qu, Fig.2Qu. Clockwise electric current $I _{p}$ can not be obtained from anticlockwise current $I _{p}$ with the help of the rotation operation of the loop in the quantum case, in contrast to the classical one, Fig.1Cl. The persistent current (4) $I _{p} = (n\Phi_{0} - BS \cos \theta)/L _{k}$, existing due to the quantization (2), will change with the loop $\bf{S}$ rotation on the angle $\theta $ relatively the $\bf{B}$ direction. The rotation operation of the loop on the angle $\pi $ can not change the $I _{p}$ direction relatively the $\bf{B}$ direction and the total energy expended during this rotation equals zero.  

The reversal of the $I _{p}$ direction is assumed due to quantum tunneling  \cite{QT1995} or quantum superposition \cite{QS2000} in the case of flux qubit and due to the transition into the normal state, Fig.1Qu, of a loop segment \cite{QF2012} or whole loop without Josephson junctions \cite{PRB2014}. The change of the quantum number from $n = 0$ to $n = 1$ changes the current direction, Fig.1Qu, as well as the turning-over of the loop with the invariable current, Fig.1Cl. According to the law of energy conservation the expenses of energy should not depend on the way of the current  reversal. This requirement was proved above in the case when magnetic field appears after the current, Fig.2Cl. It may be proved  also in the classical case when the current appears after magnetic field. The power source inducing the magnetic flux $\Phi = L _{1}I _{1} = 0.5\Phi _{0}$ with the help of the first loop should expend the additional energy 
$$\int _{t}dtI_{1}V = \int _{t}dt I_{1} L_{1,2} \frac{dI _{2}}{dt} =  \int _{t}dt I_{1} L_{1} \frac{dI _{2}}{dt} = \Phi I _{2} \eqno{(10)} $$ 
when other power source (it is not shown on Fig.2Qu) induces the current $I _{2} \ll I_{1}$ in the second loop. We can describe completely this process in the classical case using the Newton's second law $mdv/dt = qE = q(-\nabla  V - dA/dt)$ for the both loops. But we can not use the Newton's second law in the quantum case, shown on Fig.2Qu. 
  
\subsection{What is the force propelling the mobile charge carriers at quantization?}
We can describe completely using the Newton's second law the transition from $I_{2} = I _{p}$ to $I_{2} = 0$ (from left to central picture on Fig.1Qu) after switching of a loop segment in normal state. The electrical current decays exponentially from $I_{2} = I _{p}$ to $I_{2} = 0$  and the velocity of the mobile charge carriers goes down to zero under influence of the dissipation force $F _{dis}$. We can write the Newton's second law $ mdv/dt = q (-\nabla V - q dA/dt) - F _{dis} $ for mobile charge carriers in all loop segments after the transition into normal state. But we can not write it for the process (from central to right picture on the Fig.2Qu) of the $I _{p}$ appearance after the reversion of the loop segment into the superconducting state \cite{QF2012}. Quantum mechanics states that the mobile charge carriers should accelerate because of the quantization (2). But no theory says which force accelerates these carriers against the electric field force $qE = -q dA/dt$. 

This puzzle is consequence of the well-known difference between perfect conductivity and superconductivity \cite{Tinkham}. The classical case, Fig.1Cl, Fig.2Cl, in our paper corresponds to perfect conductivity, whereas the quantum case, Fig.1Qu, Fig.2Qu, corresponds to superconductivity. Behaviour of superconductor loop does not differ from the one of a perfect conductor when the wave function describing superconducting state is broken, for example with a power source, Fig.1Cl, Fig.2Cl, or normal state, Fig.1Qu, Fig.2Qu. The Meissner effect is the first and most evident experimental evidence of the difference between perfect conductivity and superconductivity. Meissner and Ochsenfeld found that not only a magnetic field is excluded from entering a superconductor, as might appear to be explained by perfect conductivity, but also that a field in an originally normal sample is expelled as it is cooled through $T _{c}$ \cite{Tinkham}. The Meissner effect is the first experimental evidence that superconductivity in contrast to perfect conductivity contradicts to Lenz's law \cite{Hirsch2007}. Jorge Hirsch wonders fairly: "{\it Strangely, the question of what is the 'force' propelling the mobile charge carriers and the ions in the superconductor to move in direction opposite to the electromagnetic force in the Meissner effect was essentially never raised nor answered to my knowledge}" \cite{Hirsch2010}. 

Cooper pairs should move in direction opposite to the electromagnetic force also in the loop shown on Fig.1Qu. Cooper pairs in an upper segment should accelerate when the lower segment marked black on the central picture of Fig.2Qu is switched from normal to superconducting state. This puzzle seems quite mysterious because the 'force' propelling the mobile charge carriers in the distant segment must be non-local \cite{QF2012}. This unknown 'force' should choose a quantum number $n$ or $n+1$ and direction in which it will accelerate Cooper pairs, i.e. left or right picture shown on Fig.1Qu. This puzzle seems especially unsolvable in the case of mechanical closing of a superconducting loop considered in the end of the paper \cite{PRB2001}. A mechanical force should act between boundaries of the Josephson junction interrupting the loop at $\Phi \neq n \Phi _{0}$ and  $\Phi \neq (n+0.5)\Phi _{0}$ because the value of the persistent current $I _{p} = I _{c}\sin (2\pi \Phi /\Phi _{0})$ and as consequence its energy (6) should depend on the gap $d$ between the boundaries. One can calculate this force using a dependence of the critical current $I _{c}$ on $d$. This force should be zero at any gap $d > 0$ when magnetic flux inside the loop $\Phi = (n+0.5)\Phi _{0}$ because the persistent current $I _{p} = I _{c}\sin (2\pi \Phi /\Phi _{0}) = I _{c}\sin 2 \pi (n+0.5) = 0$. But the persistent current (4) can not be zero because of the quantization (4) at $d = 0$. Thus, a force providing the $I _{p}$ appearance should be not only non-local but also infinite. 

\subsection{Contradiction between the theory taking into account the energy $E _{M} $ and experimental results} 
We doubt that such magical force could be introduced in any theory. At least, such force is unknown now. Therefore nobody can say what power provides the kinetic energy in the quantum case shown on Fig.2Qu, in contrast to the classical case, Fig.2Cl. The kinetic energy can not be deduced from the history of quantum current state. But the energy of magnetic dipole moment $E _{M}$ seems to be deduced from the history. The power source inducing the magnetic flux should expend an additional energy $\Phi I _{2}$ (10) because of the additional flux $\Delta \Phi _{I} = L_{1,2} I _{2}$ induced in the first loop by the current $I _{2}$ of the second loop, Fig.2Qu. Experiments testify that the persistent current induces this additional magnetic flux $\Delta \Phi _{I} $ both in superconducting loop \cite{PCScien07} and flux qubit \cite{Tanaka2002}. No theory can describe how this flux change in time from $\Delta \Phi _{I} = 0$ at $I _{2} = 0$ to $\Delta \Phi _{I} = L_{1,2} I _{p}$ at $I _{2} = I _{p}$. But this flux must change in time and induce the electric field $E = -dA/dt $ in the first loop. Therefore the power source inducing the magnetic flux should expend the additional energy $\Phi I _{p}$ (10) in order to provide zero electric field $V - d\Phi /dt = 0$ and to maintain the current $I_{1}$. Thus, our naive tendency to identify the Hamiltonian with the energy is misleading both in classical and quantum cases.

The idea of flux qubit are based on this naive tendency. Quantum superposition and quantum tunneling seem unthinkable between states with macroscopically different energy $ \epsilon = |E _{M}|$. This puzzle together with the law of angular momentum conservation \cite{QCC2010} call the idea of flux qubit in question. The naive tendency to identify the Hamiltonian with the energy at the description of quantization effects in superconductors is upheld with numerous experimental results. Most of these results can be described only if the energy $E _{M}$  of magnetic moment in an externally produced magnetic field is not taken into account. Taking this energy into account has most destroying influence on the agreement between theory and experiment in the case of weak screening. According to the universally recognized theory \cite{Tinkham} the quantum periodicity in the transition temperature \cite{LP1962} and in other parameters \cite{Letter07,toKulik2010,PCScien07,JETP07J,PerMob2001,Letter2003,PCJETP07,PLA2012} is observed at measurements of superconducting cylinder or ring because of change at $\Phi = (n'+0.5)\Phi _{0}$ of the quantum number $n$ corresponding to the minimal energy: this number $n = n'$ at $\Phi < (n'+0.5)\Phi _{0}$ and $n = n'+1$ at $\Phi > (n'+0.5)\Phi _{0}$. But the total energy $E _{tot} = E _{kin} + E _{M} = L _{k}I _{p}^{2}/2 + I _{p}\Phi = (n\Phi_{0} - \Phi )^{2}/2 L _{k} + (n\Phi_{0} - \Phi ) \Phi /L _{k} $ of the persistent current (4) should not have minimal values in these cases. The agreement between theory and experiment may be spurious. 

\section{Conclusion}
According to the law of conservation, energy of a state is the energy expended for a creation of this state. We use this definition of energy. One can not doubt that the identification of the Hamiltonian with the energy is misleading in the classical case, Fig.1Cl, Fig.2Cl, because all experimental results corroborate that the energy of magnetic moment in magnetic field exists in this case. But in the quantum case experimental results give mutually contradictory evidences. Measurements of the magnetic moment \cite{PCScien09} and magnetic flux \cite{PCScien07,Tanaka2002,PCPRL09} induced by the persistent current testify that the energy of magnetic moment in magnetic field should exist also in the quantum case. But the quantum periodicity in different parameters observed in superconductor \cite{Letter07,toKulik2010,PCScien07,JETP07J,PerMob2001,Letter2003,PCJETP07,PLA2012} and normal metal \cite{PCScien09,PCPRL09} loops can be described if only this energy is absent. This contradiction concerning the energy of magnetic moment of the persistent current both in theory and experiment demands attention and solution. No all experimental results can be described even if the Hamiltonian is identified with the energy. Some experimental results \cite{JETP07J,1Shot04,PRL06Rej,NANO2011} seem to contradict not only this naive description but even the demand of quantization (2).

\end{document}